\documentclass{desyproc}
\usepackage{enumitem}
\usepackage{color}
\RequirePackage{lineno} 

\newcommand{\Etrue}{\ensuremath{E}}
\newcommand{\Egamma}{\ensuremath{E_{\gamma}}}
\newcommand{\Ereco}{\ensuremath{E'}}
\newcommand{\Epivot}{\ensuremath{E_{0}}}
\newcommand{\probE}{\ensuremath{P_{\rm E}}}

\newcommand{\GamBkg}{\ensuremath{\Gamma_{\rm bkg}}}
\newcommand{\nSig}{\ensuremath{n_{\rm sig}}}
\newcommand{\nBkg}{\ensuremath{n_{\rm bkg}}}

\newcommand{\wSig}{\ensuremath{w_{\rm sig}}}
\newcommand{\wBkg}{\ensuremath{w_{\rm bkg}}}
\newcommand{\wROI}{\ensuremath{w^{\rm ROI}}}
\newcommand{\slocal}{\ensuremath{s_{\rm local}}}

\newcommand{\Deff}{\ensuremath{D_{\rm eff}}}
\newcommand{\Exposure}{\ensuremath{\mathcal{E}}}

\def\SPSB#1#2{\rlap{\textsuperscript{{#1}}}_{#2}}

\newcommand{\irf}[1]{\texttt{#1}}

\newcommand{\psevenrep}{\irf{P7REP}}
\newcommand{\psrc}{\irf{P7REP\_CLEAN}}

\newcommand{\fov}{{\rm Fo\kern-1ptV}}

\definecolor{orange}{rgb}{1,0.5,0}

\begin{document}

\title{Fermi-LAT and the Gamma-Ray Line Search}

\author{{\slshape Michael Gustafsson  (for the Fermi LAT collaboration)}\\[1ex]
Service de Physique Th\'eorique, Universit\'e Libre de Bruxelles, B-1050 Bruxelles, Belgium}

\contribID{Gustafsson\_Michael}

\desyproc{ULB-TH/13-11}
\acronym{Patras 2013} 

\maketitle

\mbox{ }\\
\vspace{-6.0cm}
\begin{flushright}
ULB-TH/13-11
\end{flushright}
\vspace{3.8cm}

\begin{abstract}
A distinct  signature for  dark matter in the form of weakly interacting massive particles (WIMPs) would be  the detection of a monochromatic spectral line in the gamma-ray sky. The Fermi-LAT collaboration has searched for such a line in the energy range from 5 to 300 GeV in five  sky regions around the Galactic centre. No globally significant line is detected, and 95\% CL upper limits on monochromatic-line strengths are presented.  The smallest search region reveals a line-like structure at 133 GeV with a local significance of 2.9\,$\sigma$ after 4.4 years of data, which translates to less than 1$\sigma$ global significance from a trial factor of around 200. 
\end{abstract}
\vspace{-0.2cm}

\section{Introduction}
Today, 80 years after Fritz Zwicky's first observation of a large content of invisible matter in galaxy clusters, the nature of  dark matter (DM) still remains a mystery.  By multiple observational probes it is by now well established that within the standard cosmological model about 27\% of the total energy density of the Universe is in the form of non-baryonic DM \cite{Ade:2013zuv}. From its presence at the time of the last scattering surface of the cosmic microwave background photons, DM continued to cluster and virialize into extended halos and subhalos that now host galaxy clusters and individual galaxies in their centers. The most well-studied class of models to explain the DM nature is that of weakly interacting massive particles (WIMPs). The WIMPs' interaction with standard model particles put them  in thermal and chemical equilibrium in the early Universe. As the Universe expanded and cooled down their interactions stopped, WIMPs were basically no longer diluted by pairwise annihilations, and they `froze-out', leaving a relic abundance of stable WIMPs that now constitute (cold) DM. The required effective annihilation cross-section  for WIMPs  to produce the observed DM density turns out to be of the same\footnote{Notable exceptions where the WIMP annihilation strength today and at freeze-out are different are {\it e.g.}\  when cross sections are dominated by p-wave, resonance, threshold, Sommerfeld or coannihilation processes.} 
size that can be probed by the sensitivity of current cosmic-ray telescopes. An instrument such as the Fermi Large Area Telescope (LAT) would indirectly see these WIMPs, if their mass and annihilation channels are favorable, as they  pair annihilate in high-densities of DM halos today.

One of the main challenges of detecting signals from WIMPs is that any such signal needs to be discriminated against backgrounds.
A smoking-gun signature like a monochromatic line in the cosmic gamma-ray energy-spectrum, caused by annihilation of WIMPs directly into two photons,
could  become the cornerstone for an unambiguous discovery of a DM particle signal. Recent claims ranging from tentative \cite{Bringmann:2012vr,Weniger:2012tx} to strong evidence \cite{Su:2012ft} for a gamma-ray line emission around the Galactic centre in the Fermi-LAT data have already received  hundreds of citations. The publication of a careful gamma-ray line search by the Fermi-LAT collaboration  \cite{Fermi-LAT:2013uma}  has therefore been anticipated.

\section{Analysis chain}  

\subsection{The instrument} 
The  LAT, on board the Fermi gamma-ray space telescope, is primarily a gamma-ray particle detector for energies  20 MeV to above 300 GeV \cite{Atwood:2009ez}. Incoming photons at these energies have a cross section entirely dominated by pair conversion into $e^+$$e^-$ due to quantum electrodynamical (QED) interaction with the nucleon fields in atoms. The LAT has  therefore been constructed with three detector systems to measure the cosmic gamma-ray flux: a converter/tracker system that promotes the $e^+$$e^-$ pair conversion and measures the charged particle tracks, a calorimeter composed of CsI(Tl) scintillation crystal layers adding up to 8.6 radiation length to absorb events' energy and provide an energy resolution of $\sim$10\% around 100 GeV, and finally an anticoincidence system consisting of plastic scintillator tiles that surrounds the whole tracker to reveal and reject charged cosmic-ray backgrounds  that enter the detector with up to $10^6$ larger rates than gamma-ray events.

\subsection{Event selection} 
LAT gamma-ray event reconstruction and classification algorithms have received several upgrades before and after the launch \cite{Ackermann:2012kna}. Upgrades are grouped into `\irf{Passes}',  and each such pass version includes  various event `\irf{Classes}'. These  classes,  like `\irf{Transient}' or `\irf{Clean}' ,  are designed to be optimal for different types of analysis,  and are sub-samples of events with varying degrees of gamma-ray purity;  {\it i.e.}\ more efficient cosmic-ray rejection can be provided at the price of degrading the effective area. In the years 2012 and 2013 all collected data were reprocessed using updated calibrations in the instruments reconstruction algorithms, which resulted in the pass \psevenrep\ data set  \cite{Ackermann:2012kna}. In order to prevent cosmic-ray contamination to dominate at high Galactic latitudes,  the more selective \irf{Clean} class events were used for the line search. Additional standard  quality cuts were performed  \cite{Fermi-LAT:2013uma}, and only data in the energy range 2.61-541 GeV taken from August 4, 2008 to April~4, 2012 were used for the line limit analysis and data up to December 12, 2012 in the investigation of a line-like feature around 133 GeV. It is worth noticing that the use of this reprocessed \psrc\ (version \irf{v10}, to be precise) data set differs from earlier published line-search studies.

\subsection{Regions of interest} 
Gamma-ray data from five different (but nested) sky regions of interests (ROIs) were used in the WIMP line search. These ROIs are circular regions of radius $R_\text{GC}$ around the Galactic centre with a rectangular region along the Galactic plane ($|b|<5^\circ$ and $|l|> 6^\circ$) and with known point sources \cite{Fermi-LAT:2011iqa}  (except for the smallest ROI) masked out.  These sky regions are set up to optimize the signal-to-noise ratio for different assumed DM density profiles with the signal from annihilating or decaying DM on top of a background set by the LAT team's standard model for Galactic and isotropic gamma-ray diffuse emission%
\footnote{Specifically, gal\_2yearp7v6 v0.fits and iso p7v6clean.txt, available at http://fermi.gsfc.nasa.gov/ssc/data/ access/lat/BackgroundModels.html}.
The DM distributions considered are a contracted-NFW (inner slope $\gamma=1.3$),  Einasto, NFW, and  Isothermal density profiles (as described in \cite{Fermi-LAT:2013uma}). These profiles define the ROIs: R3, R16, R41 and R90 with  $R_\text{GC}$= 3$^\circ$, 16$^\circ$, 41$^\circ$ and 90$^\circ$, respectively, for annihilating DM, and the full sky-region R180 for decaying DM.

\subsection{Model of the energy dispersion }
WIMPs in our Galaxy are primarily non-relativistic (with velocities $\lesssim 10^{-3} c$),  therefore  annihilations or decays into monochromatic photons in the center-of-mass frame become negligibly  Doppler shifted and would thus be detected as a perfect spectral line by the LAT (as its energy resolution is  $\sim$5-15\% $\gg10^{-3}$).  
The LAT deploys  two separate  energy reconstruction algorithms, and a classification-tree scheme is employed  to determine event-by-event which algorithm gave the best energy measurement.  

A notable analysis improvement ($\sim$15\% in sensitivity) compared to previous line searches \cite{Ackermann:2012qk} is achieved by employing a dispersion,  or energy probability distribution, model on an  event-by-event basis. Each event is assigned a dispersion function \Deff(\Ereco) based on its reconstruction quality $P_\text{E}$ (an event quality quantifier arising from the event reconstruction algorithms \cite{Ackermann:2012kna}). The \Deff\ can for each \Etrue\ and $P_E$ be well parametrized by a triple Gaussian function:
\begin{equation} \label{eq:TripGaus}
\Deff(\Ereco;\Etrue,\probE) = \sum_{k=1}^3 -\frac{a_k}{\sigma_k\sqrt{2\pi}}e^{-((\Ereco/\Etrue) - (1+\mu_k))^2/2\sigma_k^2}\;,
\end{equation}
with the 9 parameters $a_k$ and $\mu_k$ (constrained by $a_1+a_2+a_3=1$) determined from GEANT4/\allowbreak{}GLEAM \cite{Agostinelli:2002hh} based detector simulations.
This 2-dimensional approach (2D as it is a function of both energy $E$ and $P_E$) is particularly good when the number of signal events are low, as it  provides a better description of expected energy dispersion of the  actually observed events.

\subsection{Fitting method}
For the line search, the null hypothesis for the energy spectrum is a single power law, with exponent $\Gamma_\text{bkg}$ and flux normalization $n_{bkg}$ as free parameters, whereas the signal hypothesis adds a monochromatic line signal with free normalization \nSig.  An exposure correction $\eta(\Ereco)$ must be applied to the energy-extended background before it is fitted against observed counts data, while this is already accounted for in the effective  dispersion of the line signal. The model, including a line at energy $\Egamma$, is given by
\begin{equation}
C(\Ereco,\probE|\vec{\alpha})  
 = \nSig \Deff(\Ereco;\Egamma,\probE) \wSig(\probE)  
 + \nBkg\left(\frac{\Ereco}{\Epivot}\right)^{-\GamBkg} \eta(\Ereco) \wBkg(\probE) \,,
\end{equation}
where $\vec{\alpha}$ represent the model parameters \Egamma, \GamBkg\, \nSig\ and \nBkg.  The 2D dispersion model turns out to effectively absorb the otherwise  significant variation in \Deff\ depending on photons' incident-angles to the LAT. This, when taken together with LAT's  fairly uniform exposure of  various pointing angles of the full sky and a narrow energy band for each line fit (see below), means that the expected probability distributions $\omega(P_E)$ of $P_E$ for signal and background could be factored out and taken to be approximately equal. Moreover, with $\wSig(\probE) = \wBkg(\probE) =  \wROI(\probE)$ the otherwise subtle `Punzi effect'  \cite{Punzi:2004wh}  becomes absent.

A sliding energy-window technique is then used, where each line fit is performed in separate energy bands of $\pm 6 \sigma_\text{E}$ around \Egamma\ (where $\sigma_\text{E}$ is the energy resolution\footnote{$\sigma_E$ is defined as the half-width of the $\pm$34\% containment about the peak value at \Egamma\ of the energy  dispersion for on-axis events.}). 
The background model of a single power-law is a good approximation in such narrow energy windows. In fact, the use of narrow windows is a way to trade systematic uncertainty in the background model for larger statistical uncertainties. Energy steps of $0.5\sigma_E$ are  taken to scan over all \Egamma\ values.

For a given \Egamma, the best-fit values of \GamBkg, \nSig\ and \nBkg\ (with the restriction of \mbox{$\nSig\geq0$}) are given by maximising the following extended likelihood function for $n$ unbinned events
\begin{equation} \label{eq:L}
{\cal L} = \frac{e^{-C_\mathrm{tot}}}{n!} \prod_i^n C(\Ereco_i,P_{E_i}),
\end{equation}
where $C_\text{tot}$ is the total number of $\gamma$ rays predicted by the model. In practice, binned fits are performed for $\Egamma <25$ GeV to speed up calculations.\footnote{By the use of 60 bins, which is much narrower than the energy resolution, the information loss is negligible.}

\subsection{Statistic interpretation}
The test statistic is constructed by the maximum likelihood ratio,
\begin{equation} \label{eq:TS}
TS=2\,\textrm{ln}\frac{\mathcal{L}(\nSig = n_{\rm sig,best})}{\mathcal{L}(\nSig=0)}.
\end{equation}
The application of Chernoff's theorem \cite{Chernoff1938} predicts that the TS distribution is $\frac{1}{2} \delta(TS) + \frac{1}{2} \chi^2(TS)$, with $\chi^2$ of one degree of freedom. The local significance, in units  of standard deviations ($\sigma$), is then defined\footnote{Significance, expressed as $N$ standard deviations, will be defined by $\frac{1}{2}\int_{N^2}^{\infty} \chi^2(x)\, \text{d}x = p\text{-value}$.\label{note}} as the square-root of TS:
\begin{equation} \label{eq:slocal}
s_\text{local} = \sqrt{TS}. 
\end{equation}
%

A line is fitted at  88 different \Egamma\ values in R16, R41, R90 and R180, and 44 in R3\footnote{Energies were restricted to above 30 GeV in R3 to avoid complication due to a comparably large LAT point-spread-function and the many known point sources at the lowest energies in this ROI.}, giving a total of 396 trial fits. These trials are not independent, as ROIs are nested and energy steps not fully separated. The effective number of trials is ({\it cf.} trial factor $\equiv p_\text{global}/p_\text{local}$)
\begin{equation} \label{eq:nt}
n_t(s_x) \equiv \frac{\log(1-p_\text{global})}{\log(1-p_\text{local})},
\end{equation}
defined from the two $p$-values $p_\text{global}$ and $p_\text{local}$ that are, respectively, the probability of a {\it local} trial to have a $\sqrt{TS}$ larger than some value $s_x$, and the {\it global} ({\it i.e.}\ post all trials) probability that the maximal $\sqrt{TS}$ is larger than $s_x$. Empirically, from performing pseudo experiments on 1000 Monte Carlo generated  background simulations, it was found that an  effective number of trials $n_t \simeq 198\pm6$ gives a very good description for  conversion from local to global significance.

\section{Results}
\begin{figure}[!t]
\hspace{-0.2cm}
\includegraphics[width=0.49\textwidth]{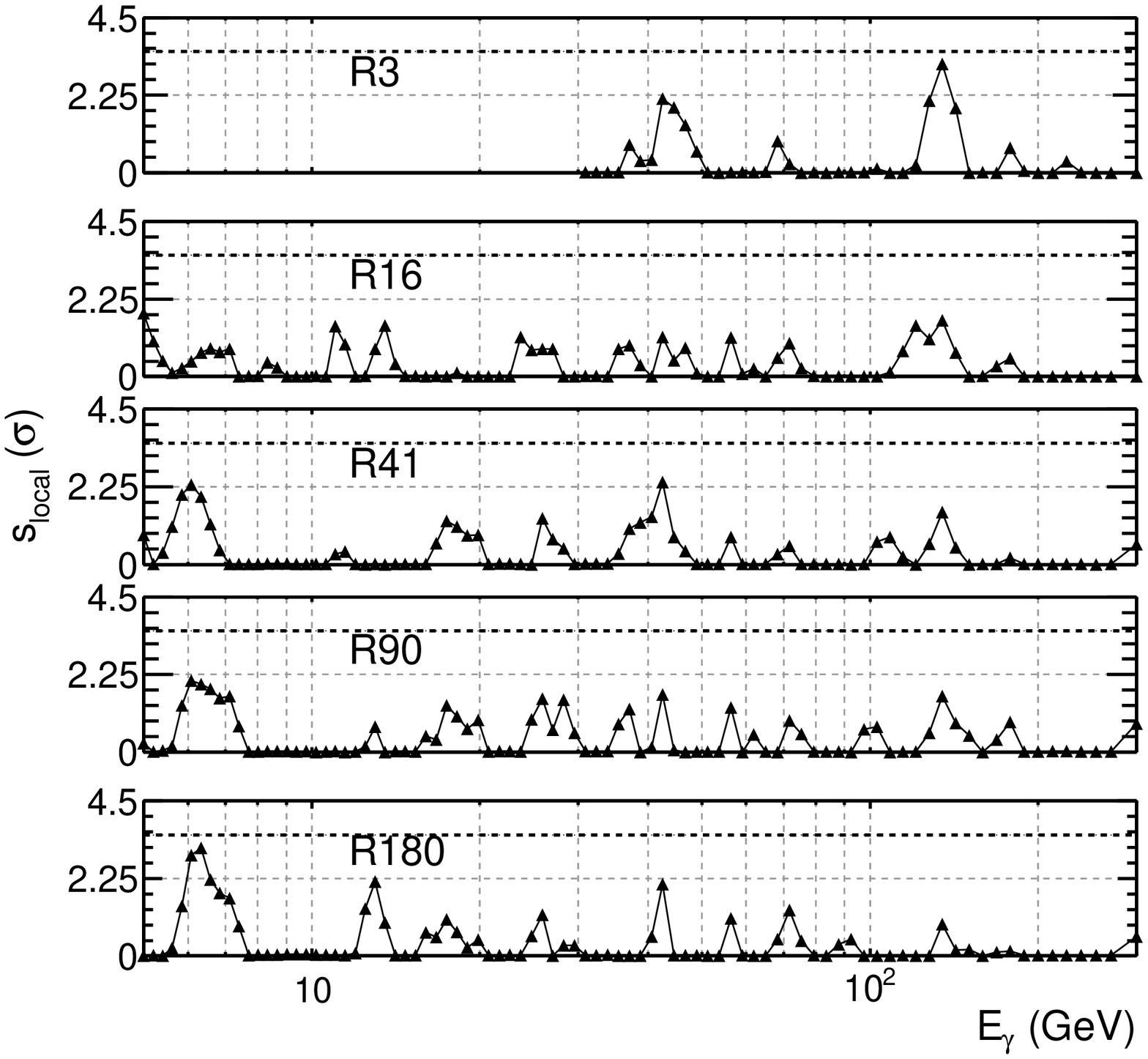} \hfill
\raisebox{0.4cm}{\includegraphics[width=0.53\textwidth]{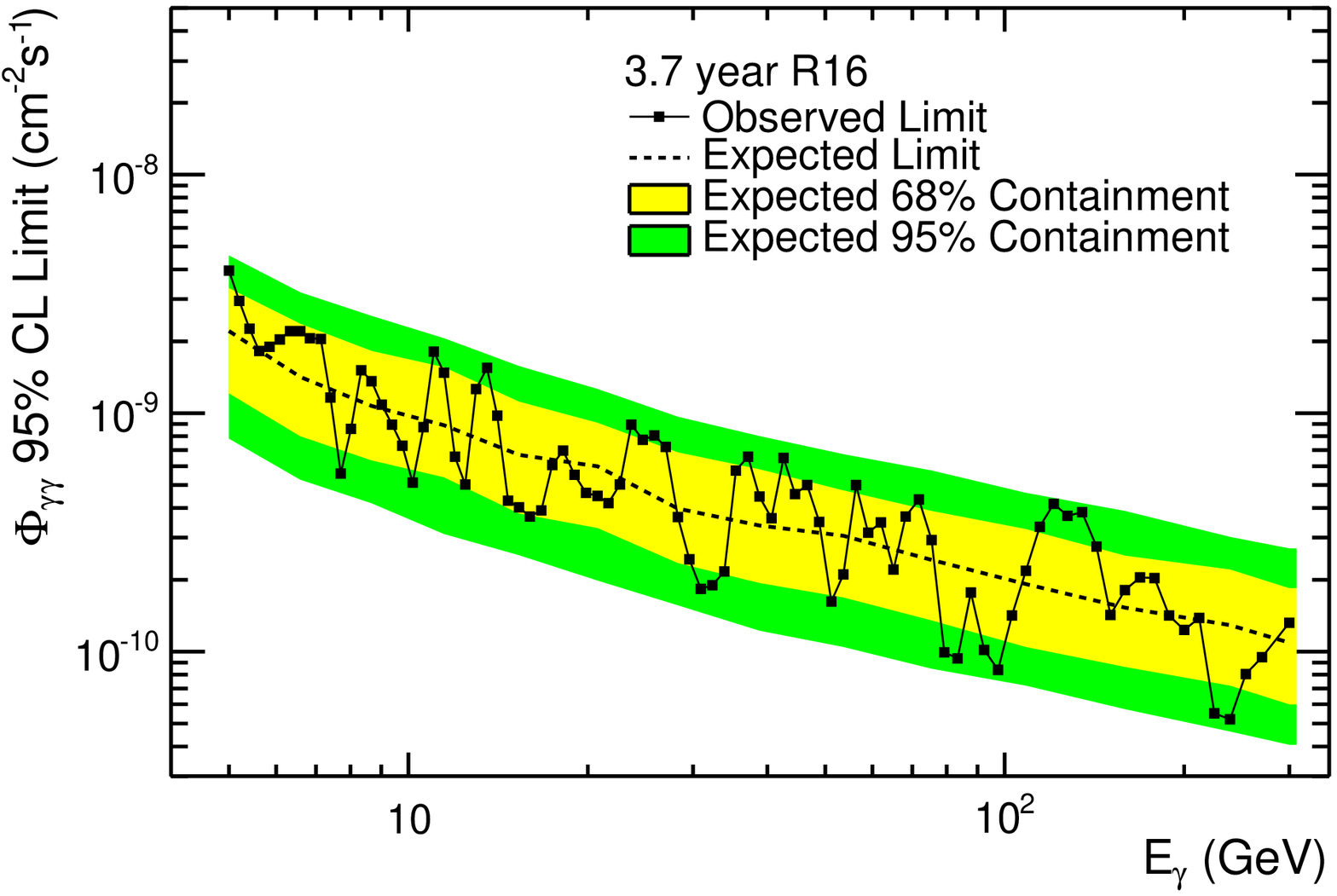}\mbox{ }}
\caption{LEFT PANEL:
Local fit significance vs. line energy in all five ROIs.  
The dashed line at the top of the plot indicates the local significance corresponding to $1.7 \sigma$ global significance\textsuperscript{\ref{note}} derived with $n_t = 198$ in Eq.\;(\ref{eq:nt}).
RIGHT PANEL: 95\% CL $\Phi_{\gamma\gamma}$ in the R16 ROI (black).  Yellow (green) bands show the 68\% (95\%) expected containment derived from 1000 single-power law (no DM) MC simulations.  The
dashed lines show the median expected limits from those simulations.}\label{fig:R16_FluxUL}
\end{figure}

The local significances, defined in Eq.~(\ref{eq:slocal}), for all tested energies and the five ROIs are shown in the left panels of Fig.~\ref{fig:R16_FluxUL}. The largest significances are found at 135 GeV in R3 and at 6.3 GeV in R180; with significances  \slocal=3.2$\sigma$ and 3.1$\sigma$, respectively.   A finer grid scan (0.1$\sigma_E$ steps) revealed that the best-fit is at 133 GeV,  with $\slocal=3.3\sigma$ and signal-to-background ratio $f_\text{R3} = 0.61\pm0.19$. However, taking into account that the effective number of trials is about 200 (or about 300 for the finer 0.1$\sigma_E$ steps)
in the search for a spectral line in the data, their global statistical significances are less than $2\sigma$.

An upper limit on $\nSig(\Egamma)$ is set at the point when the logarithm of the likelihood in  Eq.~(\ref{eq:L}) is decreased by factor 1.36 compared to its maximal value. This corresponds to a  95\% confidence-level (CL) upper limit on \nSig\ (bounded to be positive).  A limit on $\nSig$ can then be directly converted  to a 95\% CL upper limit on the line flux by 
\begin{equation} \label{eq:Phi}
\Phi_{\gamma\gamma}(\Egamma)=\frac{\nSig(\Egamma)}{\Exposure_{ROI}(\Egamma)},
\end{equation}
where $\Exposure_{ROI}$ is the LAT average exposure for the relevant ROI. The upper limits on line fluxes in region R16 are shown in the right panel of  Fig.~\ref{fig:R16_FluxUL}.

To  translate a flux limit into a velocity averaged annihilation cross section ($\langle \sigma v\rangle$) or decay life-time ($\tau$) limit requires the integrated signal contribution along the line-of-sight ($s$) and solid angle ($\Omega$) spanned by the ROI. The differential $\gamma$-ray flux  from annihilation of self-conjugated WIMPs is:
\begin{equation}\label{eq:flux_ann}
\frac{d\Phi_{\gamma\gamma}}{dE} = \frac{1}{8\pi} \, \frac{\langle \sigma v\rangle}{m_\chi^n}\,\frac{dN_\gamma}{dE} J,
\quad\text{with}\quad
J = \int  \rho_\text{DM}^n ds d\Omega,
\end{equation}
where $m_\chi$ is the WIMP mass, and $n=2$ in the exponent of the DM density ($\rho_\text{DM}$) in the `J-factor'.  Prompt annihilation into two photons yields $dN_\gamma/dE=2\delta(\Egamma-E)$ with $\Egamma = m_\chi$. Integration over the energy $E$ is trivial, and a bound on  $\Phi_{\gamma\gamma}$ directly translates  to a $\langle \sigma v\rangle$ bound for a given $J$ value.
%
\begin{figure}[!t]
\includegraphics[width=0.495\textwidth]{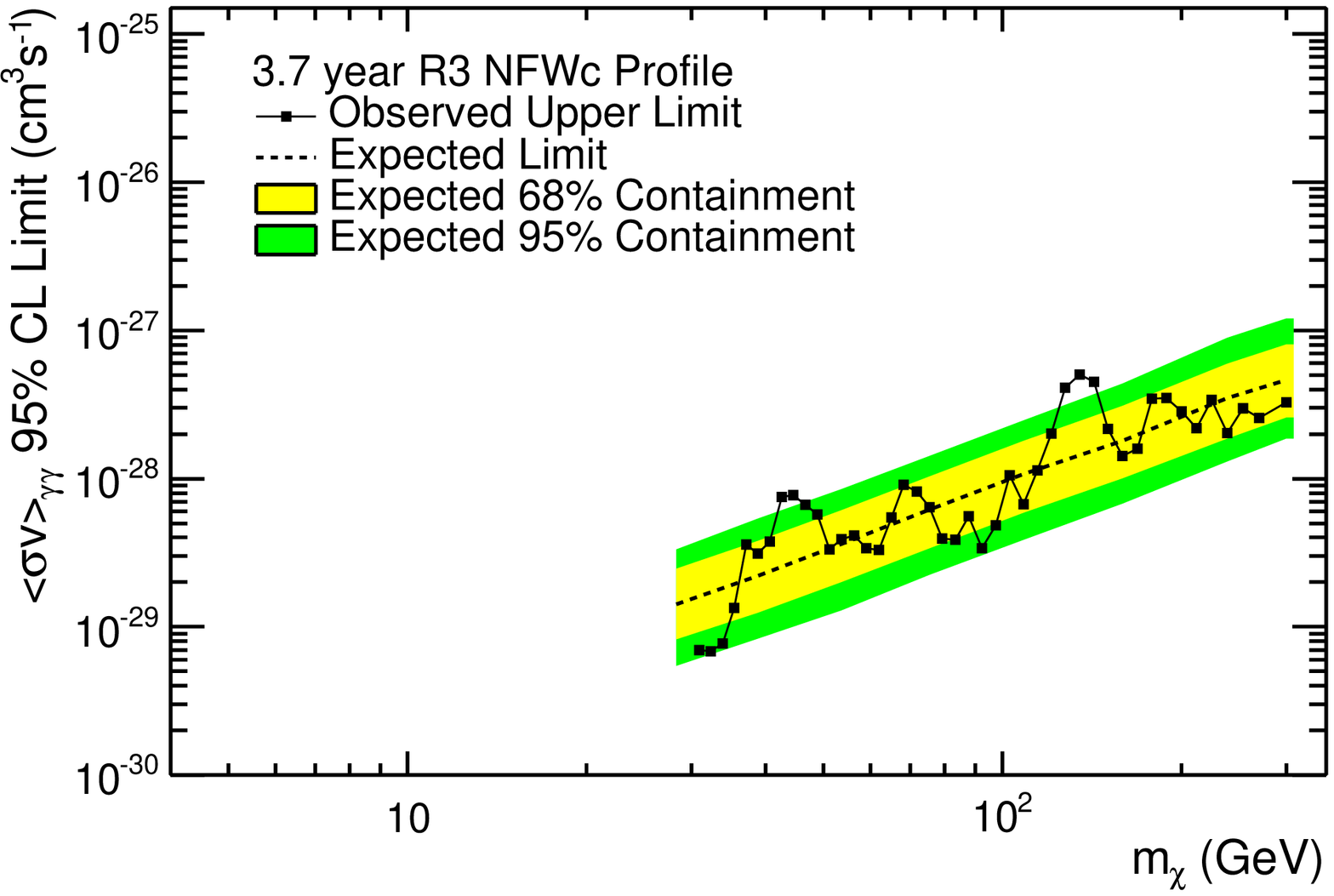} \hfill
\includegraphics[width=0.495\textwidth]{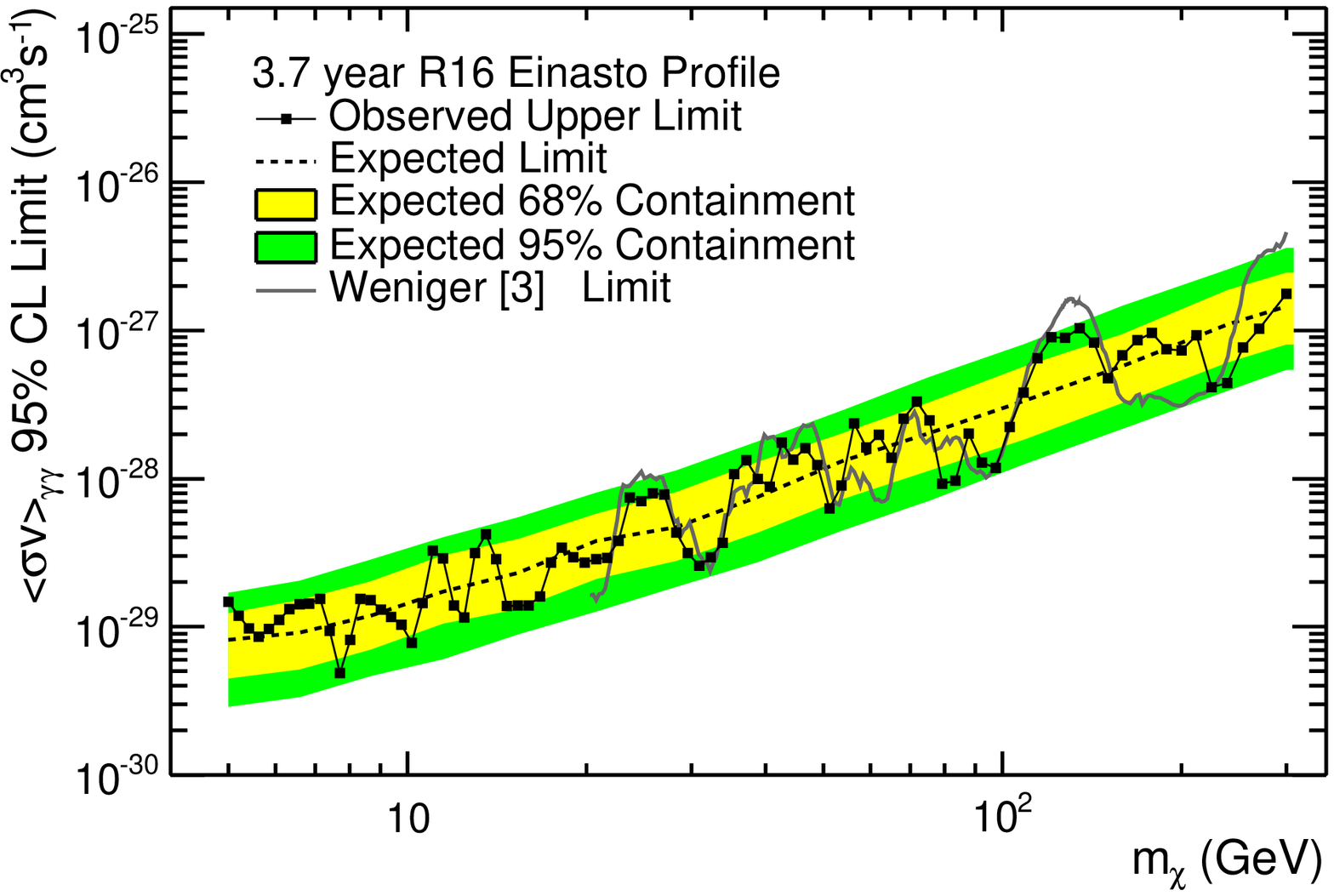}
\caption{95\% CL $\langle\sigma v\rangle_{\gamma\gamma}$ upper limits for our contracted NFW  
and Einasto  DM profile in ROI R3  and R16, respectively.  Colored bands and lines are defined as in Fig.~\ref{fig:R16_FluxUL}. The solid gray line shows the limits derived in \cite{Weniger:2012tx} for a comparable ROI and identical DM density profile.}\label{fig:SigmaVUL}
\end{figure}
Figure~\ref{fig:SigmaVUL} shows the limits from the contracted NFW profile and the Einato profile; with  $J_\text{R3}=13.9\cdot10^{22}$~GeV$^2$cm$^{-5}$ and $J_\text{R16}=8.48\cdot10^{22}$~GeV$^2$cm$^{-5}$ in their optimized ROI.
In the case of decaying DM into \emph{one} monochromatic photon, perform the following replacements $\langle \sigma v\rangle \rightarrow 1/\tau$, $n\rightarrow1$, $m_\chi \rightarrow m_\chi/2$, and  $dN_\gamma/dE=\delta(m_\chi/2-E)$ to calculate $\Phi_{\gamma\gamma}$.

\subsection{Systematic uncertainties}
Systematic uncertainties are not included in the presented limits, therefore it is important to investigate how much this could  alter the results.  Table \ref{tab:Syst_Error_By_ROI} summaries the potential size of such effects, where uncertainties have been split into three classes \cite{Fermi-LAT:2013uma}:\\
\vspace{-0.8cm}
%
\begin{wraptable}{r}{0.46\textwidth}
\begin{tabular}{|clcc|}
\hline
Quantity 				& Energy 			& R3/R16 					& R180 \\
\hline
$\delta f$ 				& 5~GeV     		& $\pm$0.020 		 		&  $\pm$0.008 \\
$\delta f$ 				& 50~GeV   		& $\pm$0.024 		 		& $\pm$0.015 \\
$\delta f$ 				& 300~GeV 		& $\pm$0.032 		 		&  $\pm$0.035 \\
$\delta \nSig / \nSig$  	& All				 & $\SPSB{+0.07} {-0.12}$     	& $\SPSB{+0.07}{-0.12}$ \\
$\delta\Exposure / \Exposure $ & 5~GeV 		& $\pm$0.10 				&  $\pm$0.14 \\
$\delta\Exposure / \Exposure $ & 300~GeV 	& $\pm$0.10 		 		&  $\pm$0.16 \\
\hline
\end{tabular}
\caption{\label{tab:Syst_Error_By_ROI} Magnitude of systematic effects, by ROI and Energy, where all contributing uncertainties have been added in quadrature.}
\end{wraptable}
\begin{enumerate}[leftmargin=0.4cm,itemindent=0.0cm,labelwidth=\itemindent,labelsep=4pt,itemsep=-0.2pt]
\item[1.] Those that  would induce a false signal or mask a true signal $\delta f$; from  {\it e.g.}\  unmodeled  variation in the effective area, imperfect background model and cosmic-ray contamination. 
\item[2.] Those that would scale the fit estimates of  the number of signal counts $\delta\nSig$; from {\it e.g.}\ imperfect line dispersion model and $\Egamma$ grid spacing.   
\item[3.]  Signal to flux conversion uncertainties $\delta\Exposure$; from exposure and effective area uncertainties. 
\end{enumerate}
%
\mbox{}\\ \\ \vspace{-0.9cm}

One can now see that the systematic uncertainty $\delta f$ in R180 at low energies is of the order 1\%, which therefore could be an explanation for the  $3.1\sigma$ signal significance seen at  6.3 GeV that had $f=0.010\pm0.002$.  The signal fraction for the 135 GeV feature is, on the other hand,  $f=0.58\pm0.18$ in R3 and thus much larger than known systematic effects.
Notably, a  spectral structure at 133 GeV with a significance of $2.0\sigma$ is  also found in Earth albedo/limb data, which is used as a  (DM free)  bright gamma-ray control-region. This could indicate an instrumental issue at 133 GeV, but due to its much smaller signal-to-background fraction $f=0.14\pm0.07$ and no line-like structure detected in a second control region, the Galactic plane, the line-like signal around the Galactic center cannot unambiguously be attributed to any instrumental effect.


\subsection{Line-like feature at 133 GeV}
The most significant fit from the line search was at \Egamma=135 GeV in the  R3 search region. Several steps in investigating the significance and properties of this spectral structure have been made. The impact of  various changes in datasets and modeling the signal can be summarized as:
\begin{itemize}[leftmargin=0.6cm,itemindent=0.0cm,labelwidth=\itemindent,labelsep=6pt,itemsep=-5pt]
\item  3.7 years \irf{P7\_CLEAN} (un-reprocessed) data with a 1D dispersion (no use of $P_E$) line fit gives: 
$\slocal=4.5\sigma$ at 130 GeV.   
\item  3.7 years \psrc\ (reprocessed) data with a 1D dispersion (no use of $P_E$) line fit gives: 
$\slocal=4.1\sigma$ at 133 GeV.   
\item  3.7 years \psrc\ (reprocessed) data with a 2D dispersion (including $P_E$) line fit gives: 
$\slocal=3.3\sigma$ at 133 GeV.   
\item  4.4 years \psrc\ (reprocessed) data with a 2D dispersion (including $P_E$) line fit gives: 
$\slocal=2.9\sigma$ at 133 GeV.   
\end{itemize}
\begin{figure}[!tb]
\center{\includegraphics[width=0.55\textwidth]{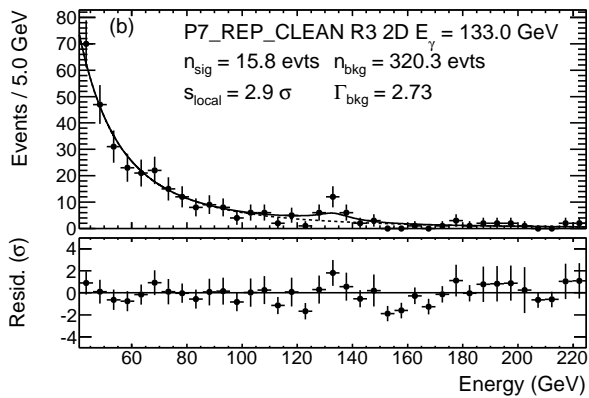}}
\caption{Fit for a line-signal signal at 133~GeV in R3 using a 4.4 year \psrc\ dataset with the 2D energy dispersion model. The solid curve shows the average model weighted using the $P_E$ distribution of the fitted events.  The data binning is only for visualization purposes.}\label{fig:133_4.4yr} 
\end{figure}
The first point allows a good comparison to previous works and, even if {\it e.g.}\ our ROI differ somewhat from others works \cite{Weniger:2012tx}, it is clear that a local significance of above $4\sigma$ can be reached with this setup. 
In the next point, when reprocessed data is analyzed, the significance becomes about 10\% smaller. Gamma-ray events  might switch \irf{classes} \cite{Bregeon:2013qba} due to reprocessed  data, and it is only about 70-80\%  overlap of events in \irf{P7\_CLEAN}  and \psrc. The shift of \Egamma\ of 130 to 133 GeV is due to the correction of a known $\sim$1\% per year degrading in light yield efficiency in the calorimeter crystals.
The next point incorporates the switch to the 2D dispersion model given in Eq.~(\ref{eq:TripGaus}), which decreases the significance by around 20\%. The actual events around 133 GeV have lower $P_E$  on average than expected, which cause the dispersion function to broaden and lower the likelihood value for events close to 133 GeV. In Fig.~\ref{fig:133_4.4yr} it is clearly visible how the line-like feature seems narrower than the expected LAT energy  dispersion. 
To quantify this, a rescaled energy width of our 2D dispersion was tested
. The best-fit result was to rescale the width by a factor $s_\sigma=0.32^{+0.11}_{-0.07}$, which also increased $TS$ by 9.4. Pre-launch beam tests show however that the width is known within 10\% up to 280 GeV, and this $\sim$0.32 times narrower dispersion functions is thus inconsistent with the pre-measured shape at the 2-3$\sigma$ level.  
Finally, the last point includes more data (extending from 3.7 to 4.4 years) which further reduces the significance by about 10\%, to a local significance of 2.9$\sigma$. 

It is worth mentioning that the actual amount of change in significance depends on the specific ROI. For example, for Weniger's analysis in his Region 3 \cite{Weniger:2012tx}, the impact of going from \irf{P7\_CLEAN} to \psrc\ dilutes the original significance of 4.3$\sigma$ to $2.8\sigma$, and then by 
utilizing a  2D dispersion model to $2.4\sigma$, and finally by also extending to 4.4 years of data diminishes the local significance furter to $2.0\sigma$ \cite{Weniger}. 
%
All these additional tests around \Egamma=133 GeV are outside the original line search, so their contribution to the trials factor is difficult to estimate precisely. It is however clear that taking the 2D dispersion model, applied to the 4.4 years reprocessed data-set, gives a 2.9$\sigma$ local significance, which is less than a $1\sigma$ global significance when a trail factor of $\sim$200 is applied.

\section{Summary and outlook}
The hint of a line signal around $130$ GeV towards the Galactic centre has resulted in enormous interest, as it could be a long-sought WIMP DM signal.  The Fermi-LAT Collaboration has searched for spectral lines from 5-300 GeV in five ROIs without finding any globally significant lines. A line-like feature at 133 GeV with a  signal-to-noise fraction larger than known systematic effects is present. However, with reprocessed \irf{P7REP\_CLEAN} data, improved 2D line dispersion model and including 4.4 years of data the local significance is 2.9$\sigma$, which translate to a global significance below 1$\sigma$. A statistical fluctuation is therefore a possible explanation. 
It will be very interesting to re-perform a line search with Fermi-LAT's upcoming \irf{Pass 8}  data set \cite{Atwood:2013rka}, which will provide a larger effective area and the benefit of having almost all event reconstruction algorithms rewritten. 
The Fermi users' group has also already endorsed a recommendation that the Fermi mission undertake a new observing strategy that emphasizes coverage of the Galactic center region from December 2013 -- which would lead to, on average, a doubled exposure rate of the region around the Galactic centre compared with the currently used survey mode.

\bigskip
\begin{footnotesize}
\noindent
M.G. is supported by the Belgian Science Policy (IAP VII/37), the IISN and the ARC project.
The $Fermi$ LAT Collaboration acknowledges support from a number of agencies and institutes.
These include NASA and DOE in the United States, CEA/Irfu and IN2P3/CNRS in France, ASI and INFN in Italy, MEXT, KEK, and JAXA in Japan, and the K.~A.~Wallenberg Foundation, the Swedish Research Council and the National Space Board in Sweden. Additional support from INAF in Italy and CNES in France for science analysis during the operations phase is also gratefully acknowledged.



\end{footnotesize}

\end{document}